\pdfoutput=1
\documentclass[12pt]{article}
\usepackage[hiresbb]{graphicx}
\usepackage{wrapfig}
\usepackage{amsmath,amsthm,amssymb}
\usepackage{enumerate}
\usepackage{cases}
\usepackage{bm}
\usepackage{cite}

\title{Vortex solutions of the generalized Beltrami flows to the Euler equations}
\author{}

\begin{document}
\rightline{January 2015}
\rightline{~~~~~~~}
\vskip 1cm
\centerline{\large \bf Vortex solutions of the generalized Beltrami flows}
\centerline{\large \bf to the incompressible Euler equations}
\vskip 1cm

\centerline
{{Minoru Fujimoto${}^1$
 }, 
 {Kunihiko Uehara${}^2$
 } and 
 {Shinichiro Yanase${}^3$
 }
}
\vskip 1cm
\centerline{\it ${}^1$Seika Science Research Laboratory,
Seika-cho, Kyoto 619-0237, Japan}
\centerline{\it ${}^2$Department of Physics, Tezukayama University,
Nara 631-8501, Japan}
\centerline{\it ${}^3$Graduate School of Natural Science and Technology, Okayama University,
Okayama 700-8530, Japan}
\vskip 1cm
\centerline{\bf Abstract}
\vskip 0.2in
  As for the solutions of the generalized Beltrami flows 
to the incompressible Euler equations besides the solutions 
separating radius and axial components, 
there are only several solutions found 
as the Hill's vortex solutions. 
We will present a series of vortex solutions in this category
for the generalized Beltrami flows to the incompressible Euler equations.

\vskip 5mm
%
\noindent
PACS number(s): 47.10.ad, 47.15.ki, 47.32.-y
\vskip 5mm

\setcounter{equation}{0}
\addtocounter{section}{0}
\hspace{\parindent}

  It is a well-known fact for many years 
that the Hill's spherical vortex \cite{Hill} is one of exact solutions 
to the incompressible Euler equations. 
A characteristic feature of the solution is that 
the vorticity only exists inside the sphere, 
where the vorticity $\boldsymbol{\omega}$ is referred by the velocity $\boldsymbol{u}$ as
\begin{equation}
  \boldsymbol{\omega}=\boldsymbol{\nabla}\times\boldsymbol{u}.
\end{equation}

  From a point of view in the differential equations, 
the Hill's solution belongs to the solutions of 
the generalized Beltrami flows 
\begin{equation}
  \boldsymbol{\nabla}\times(\boldsymbol{\omega}\times\boldsymbol{u})=0
\label{e002}
\end{equation}
to the incompressible Euler equations. 
One category \cite{Berker,Terrill,Wang2,Weinbaum} is made of the solutions 
separating radius and axial components in the cylindrical coordinates 
where the solutions are described by the Bessel and exponential functions, 
the other is of the solutions as the Hill's vortex solution
which are non-separable as for radius and axial components 
where there are only several solutions found so far. 

  The incompressible Euler equations
\begin{eqnarray}
  \frac{\partial \boldsymbol{u}}{\partial t}
  +(\boldsymbol{u}\cdot\nabla)\boldsymbol{u}&=&-\frac{1}{\rho}\nabla p,
\label{e003}\\
  \nabla\cdot\boldsymbol{u}&=&0,
\label{e004}
\end{eqnarray}
where $p$ is the pressure and $\rho$ is the density, 
are the non-linear differential equations because of 
the existence of the second term of the left-hand side in (\ref{e003}). 
The generalized Beltrami condition (\ref{e002}) is 
a linearization condition for the equations. 
This is seen by taking the curl of (\ref{e003}) as 
\begin{equation}
  \frac{\partial \boldsymbol{\omega}}{\partial t}
  +\nabla\times(\boldsymbol{\omega}\times\boldsymbol{u})=0.
\label{e005}
\end{equation}

  As for {\it the exact solutions} to the incompressible Euler equations 
in this category, namely non-separable solutions, 
the Agrawal's\footnote{$\psi=a_2\eta^2z^2$ in (\ref{e010}) 
is a special case of the Agrawal solution, but this is a separable case 
when other coefficient are all zero.}\cite{Agrawal}, 
the Berker's \cite{Berker}, the O'Brien's \cite{OBrien} and the Wang's solution \cite{Wang} are known 
besides the Hill's vortex solution.
These solutions are all axisymmetric flows which demand the conditions
\begin{eqnarray}
  \boldsymbol{u}(\eta,\varphi,z)&=&(u_\eta,0,u_z),
\label{e006}\\
  \boldsymbol{\omega}(\eta,\varphi,z)&=&(0,\omega_\varphi,0),
\label{e007}
\end{eqnarray}
where we take the cylindrical coordinate $(\eta,\varphi,z)$. 
When we introduce the Stokes stream function $\psi(\eta,\varphi,z)$, 
which automatically satisfies the continuity condition (\ref{e004}), 
the equations (\ref{e003}) reduce to 
\begin{equation}
  \frac{\partial}{\partial \eta}\left(\frac{1}{\eta}\frac{\partial\psi}{\partial\eta}\right)
  +\frac{1}{\eta}\frac{\partial^2\psi}{\partial z^2}=-\omega_\varphi.
\label{e008}
\end{equation}

Meanwhile Marris and Aswani\cite{Marris} demonstrated 
that the only solution for these axisymmetric generalized Beltrami flows is 
that $\omega_\varphi$ is proportional to $\eta$. 
Then (\ref{e008}) becomes 
\begin{equation}
  \frac{\partial^2\psi}{\partial\eta^2}
  -\frac{1}{\eta}\frac{\partial\psi}{\partial\eta}
  +\frac{\partial^2\psi}{\partial z^2}=-\alpha\eta^2,
\label{e009}
\end{equation}
where $\alpha$ is constant. 
There is a description ''A simple, but useful, set of solutions''\cite{Drazin} 
of (\ref{e009}), which is
\begin{equation}
  \psi=a_1\eta^4+a_2\eta^2z^2+a_3\eta^2+a_4\eta^2z
       +a_5(\eta^6-12\eta^4z^2+8\eta^2z^4),
\label{e010}
\end{equation}
where $8a_1+2a_2=-\alpha$, but $a_i$'s are otherwize arbitrary. 
In order to genarate the right-hand side of (\ref{e009}), 
we have to put the term of $\eta^4$ or $\eta^2z^2$ in $\psi$, 
where {\it the exact solutions} above are all contained 
the either term or the both terms. 
In other words, the third,the forth or the fifth term of the left-hand side 
in (\ref{e010}) where the each coefficient $a_3$,$a_4$ or $a_5$ is arbitray, 
satisfies that the left-hand side of (\ref{e009}) 
equals zero. 

  The extention of the solution space for (\ref{e009}) is 
to add higher power terms for $\eta$ and $z$ which 
satisfy the same condition as $a_3,a_4$ or $a_5$ term. 
We present the result, a series of higher power terms, as
\begin{eqnarray}
  \psi&=&a_1\eta^4+a_2\eta^2z^2+a_3\eta^2+a_4\eta^2z
\nonumber\\
      &&+c_n\sum_{k=0}^{n-1}a_{n,k}\eta^{2n-2k}z^{2k}, 
\label{e011}
\end{eqnarray}
where $n\ge 3$,\ $a_{n,0}=1$ and 
$\displaystyle{a_{n,k}=\frac{(-4)^k\varGamma(n)\varGamma(n+1)}{\varGamma(2k+1)\varGamma(n-k)\varGamma(n-k+1)}}$.

\vspace{3mm}
  When the parameters are taken appropriately as 
$a_1,a_3,c_3\ne 0$ and other coefficients vanish in (\ref{e011}), 
we reproduce the Wang's toroidal vortex shown in Fig.1.

\begin{figure}[htb]
 \centering
  \includegraphics[width=0.6\columnwidth,clip]{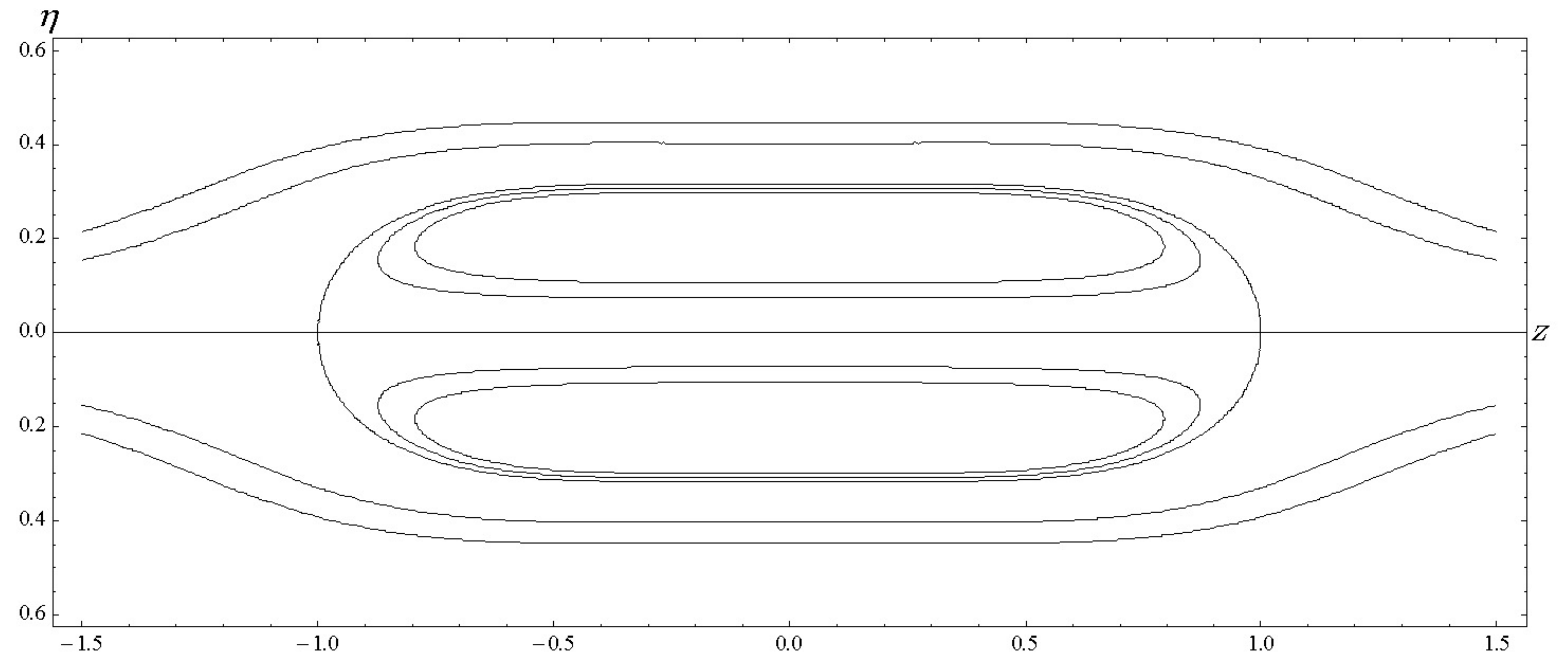}
  \caption{\small $a_1=10,a_3=-1,c_3=1/8$. The streamsurfaces have 
$\psi=-0.005,-0.01,0,0.1,0.2$ from the innermost to the outer side.}
\end{figure}%

  The examples for higher order solutions are shown in Fig.2 and Fig.3.
Their stream functions are
\begin{equation}
  \psi=a_1\eta^4+a_3\eta^2
      +c_4\left(\eta^8-24\eta^6z^2+48\eta^4z^4-\frac{64}{5}\eta^2z^6\right)
\label{e012}
\end{equation}
and 
\begin{equation}
  \psi=a_1\eta^4+a_3\eta^2
      +c_5\left(\eta^{10}-40\eta^8z^2+160\eta^6z^4-128\eta^4z^6+\frac{128}{7}\eta^2z^8\right)
\label{e013}
\end{equation}
respectively. 

\begin{figure}[htb]
 \centering
  \includegraphics[width=0.6\columnwidth,clip]{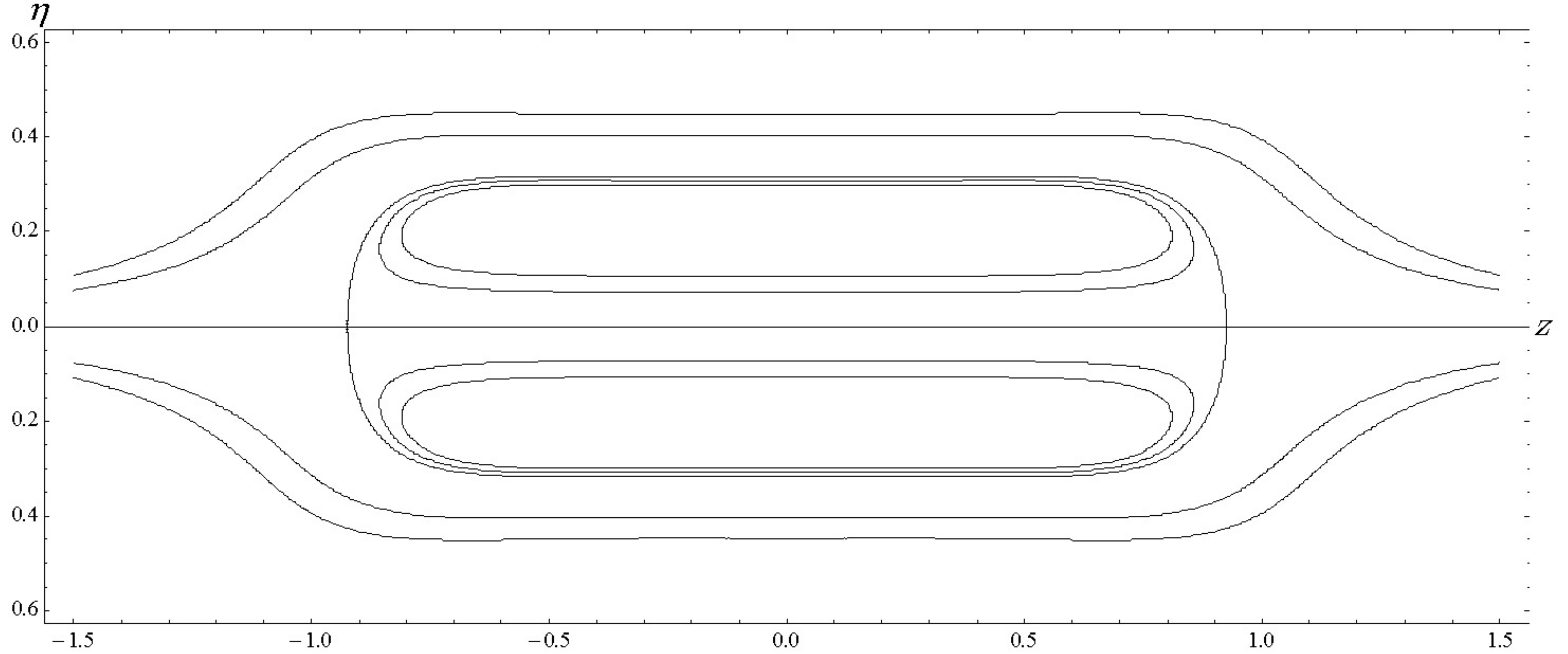}
  \caption{\small $a_1=10,a_3=-1,c_4=-1/8$. The streamsurfaces have 
$\psi=-0.005,-0.01,0,0.1,0.2$ from the innermost to the outer side.}
\end{figure}%

\begin{figure}[htb]
 \centering
  \includegraphics[width=0.6\columnwidth,clip]{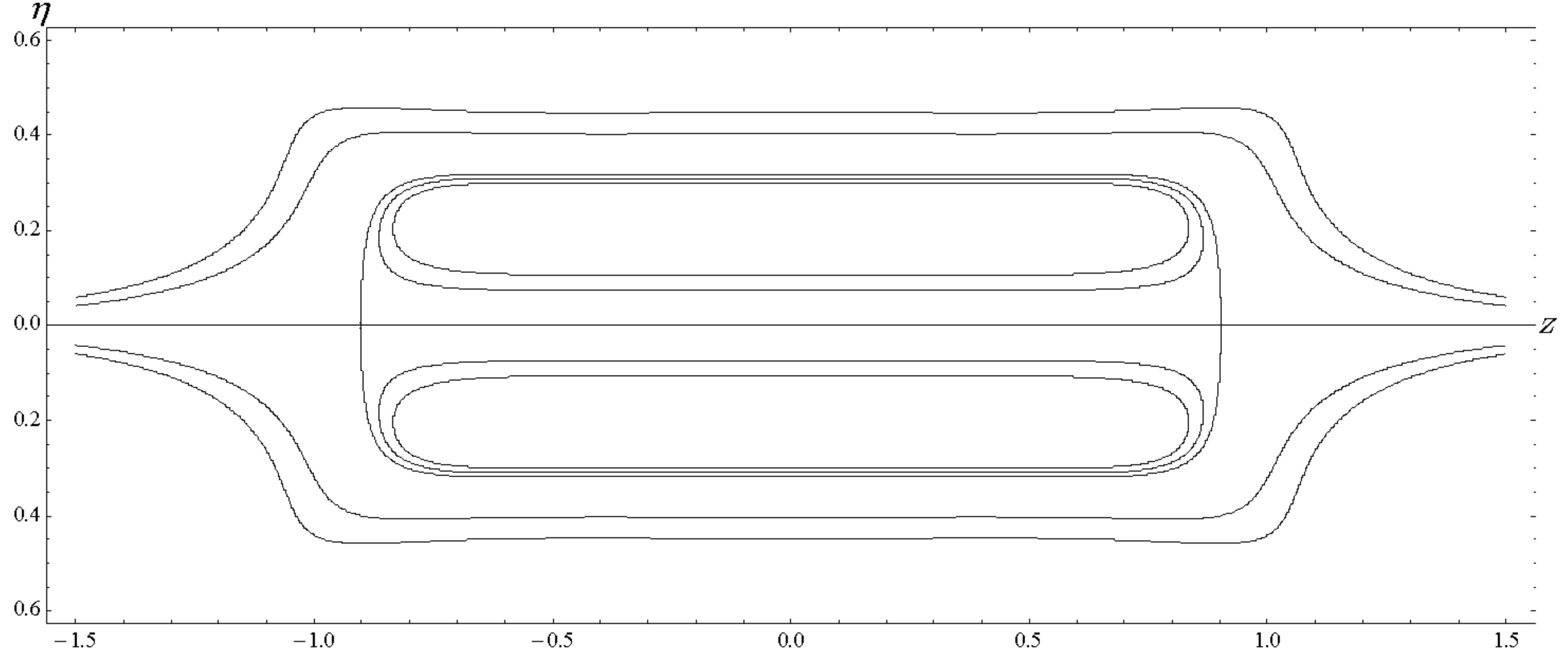}
  \caption{\small $a_1=10,a_3=-1,c_5=1/16$. The streamsurfaces have 
$\psi=-0.005,-0.01,0,0.1,0.2$ from the innermost to the outer side.}
\end{figure}%
\vspace{3mm}
  Although we have to tune the parameters for a vortex 
like in the figures to appear in each orders, 
the higher order of $\eta$ and $z$ solutions we take, 
we get the streamsurface of $\psi=0$ in the shape of 
from an oval as the Wang's vortex 
to more cylinder-like boundaries by the same order values of parameters. 
The toroidal vorticity $\omega_\varphi=\alpha\eta$ in each case though, 
these vorticity solutions should be the solutions 
inside the streamsurface of $\psi=0$. 
The reason is that the global vorticity often generates some disturbances 
in the streamsurfaces and 
this will be easily confirmed when we draw the figures 
in larger ranges of $\eta$ and $z$. 

  So, the remaining work we have to do is 
that we pour the perfect fluid outside of the boundaries of $\psi=0$ 
and connect these solutions to the outer solutions with no-vorticity 
like the Hill's solution. 

\vskip 5mm

\end{document}